\documentclass[12pt]{article}
\usepackage{axodraw,bbold}

\catcode`@=12
\begin{document}
\thispagestyle{empty}
\begin{flushright} 
UCRHEP-T442\\ 
October 2007\
\end{flushright}
\vspace{0.5in}
\begin{center}
{\LARGE	\bf A$_4$ Symmetry and Neutrinos\\}
\vspace{1.5in}
{\bf Ernest Ma\\}
\vspace{0.2in}
{\sl Department of Physics and Astronomy, University of California,\\}
\vspace{0.1in}
{\sl Riverside, California 92521, USA\\}
\vspace{1.5in}
\end{center}

\begin{abstract}\
I recount briefly the history of neutrino tribimaximal mixing and the use of 
the discrete family symmetry A$_4$ in obtaining it.
\end{abstract}
\vskip 1.0in
\noindent Talk at the 4th International Conference on Flavor Physics, Beijing, 
September 2007.

\newpage
\baselineskip 24pt

\section{Neutrino Tribimaximal Mixing}	

In 1978, soon after the putative discovery of the third family of leptons and 
quarks, it was conjectured by Cabibbo \cite{c78} and Wolfenstein \cite{w78} 
independently that
\begin{equation}
U_{l \nu}^{CW} = \frac{1}{\sqrt{3}} \left( \begin{array}{ccc} 1 & 1 & 1 \\ 
1 & \omega & \omega^2 \\  1 & \omega^2 & \omega \end{array} \right),
\end{equation}
where $\omega = \exp(2 \pi i/3) = -1/2 + i \sqrt{3}/2$. This should dispel 
the \underline{myth} that everybody expected small mixing angles in the 
lepton sector as in the quark sector.

In 2002, after much neutrino oscillation data have been established, 
Harrison, Perkins, and Scott \cite{hps02} proposed the tribimaximal mixing 
matrix, i.e.
\begin{equation}
U_{l \nu}^{HPS} = \left( \begin{array}{ccc} \sqrt{2/3} & 1/\sqrt{3} & 0 \\ 
-1/\sqrt{6} & 1/\sqrt{3} & -1/\sqrt{2} \\ -1/\sqrt{6} & 1/\sqrt{3} & 
1/\sqrt{2} \end{array} \right) \sim (\eta_8, \eta_1, \pi^0),
\end{equation}
where the 3 columns are reminiscent of the meson nonet.

In 2004, I discovered \cite{m04} the simple connection:
\begin{equation}
U_{l \nu}^{HPS} = (U_{l \nu}^{CW})^\dagger \left( \begin{array}{ccc} 
1 & 0 & 0 \\ 0 & 1/\sqrt{2} & -1/\sqrt{2} \\ 0 & 1/\sqrt{2} & 1/\sqrt{2} 
\end{array} \right) \left( \begin{array}{ccc} 0 & 1 & 0 \\ 1 & 0 & 0 \\ 
0 & 0 & i \end{array} \right).
\end{equation}
This means that if
\begin{equation}
{\cal M}_l = U_{l \nu}^{CW} \left( \begin{array}{ccc} m_e & 0 & 0 \\ 
0 & m_\mu & 0 \\ 0 & 0 & m_\tau \end{array} \right) (U_R^l)^\dagger
\end{equation}
and ${\cal M}_\nu$ has $2-3$ reflection symmetry, with zero $1-2$ and $1-3$ 
mixing, i.e
\begin{equation}
{\cal M}_\nu = \left( \begin{array}{ccc} a+2b & 0 & 0 \\ 0 & a-b & d \\ 
0 & d & a-b \end{array} \right),
\end{equation}
$U_{l \nu}^{HPS}$ will be obtained, but \underline{how}? Tribimaximal mixing 
means that
\begin{equation}
\theta_{13}=0, ~~~ \sin^2 2 \theta_{23} = 1, ~~~ \tan^2 \theta_{12} = 1/2.
\end{equation}
In 2002 (when HPS proposed it), world data were not precise enough to test 
this idea.  In 2004 (when I derived it), SNO data implied $\tan^2 \theta_{12} 
= 0.40 \pm 0.05$, which was not so encouraging.  Then in 2005, revised SNO 
data obtained $\tan^2 \theta_{12} = 0.45 \pm 0.05$, and tribimaximal mixing 
became a household word, unleashing a glut of papers.

\section{Tetrahedral Symmetry A$_4$}

For 3 families, we should look for a group with a \underline{3} 
representation, the simplest of which is A$_4$, the group of the even 
permutation of 4 objects.  It has 12 elements, divided into 4 equivalence 
classes, and 4 irreducible representations: \underline{1}, \underline{1}$'$, 
\underline{1}$''$, and \underline{3}, with the multiplication rule
\begin{eqnarray}
\underline{3} \times \underline{3} &=& \underline{1}~(11+22+33) + 
\underline{1}'~(11+\omega^2 22 + \omega 33) + \underline{1}''~(11 + 
\omega 22 + \omega^2 33) \nonumber \\ &+& \underline{3}~(23,31,12) 
+\underline{3}~(32,13,21).
\end{eqnarray}
A$_4$ is also the symmetry group of the regular tetrahedron, one of the 5 
perfect geometric solids in 3 dimensions and identified by Plato as ``fire'' 
\cite{m02}.  It is a subgroup of both SO(3) and SU(3).  The latter also 
has 2 sequences of finite subgroups which are of interest: $\Delta(3n^2)$ 
has $\Delta(12) \equiv$ A$_4$ and $\Delta(27)$; $\Delta(3n^2-3)$ has 
$\Delta(24) \equiv$ S$_4$.

\section{How 1. is obtained by 2.}

There are 2 ways to achieve Eq.~(4).  The original proposal \cite{mr01,bmv03} 
is to assign $(\nu_i,l_i) \sim \underline{3}, ~l^c_i \sim \underline{1}, 
\underline{1}', \underline{1}''$, then with $(\phi^0_i,\phi^-_i) \sim 
\underline{3}$,
\begin{equation}
{\cal M}_l = \left( \begin{array}{ccc} h_1 v_1 & h_2 v_1 & h_3 v_1 \\ 
h_1 v_2 & h_2 \omega v_2 & h_3 \omega^2 v_2 \\ 
h_1 v_3 & h_2 \omega^2 v_3 & h_3 \omega v_3 \end{array} \right)
= \left( \begin{array}{ccc} 1 & 1 & 1 \\ 1 & \omega & \omega^2 \\ 
1 & \omega^2 & \omega \end{array} \right) \left( \begin{array}{ccc} 
h_1 v & 0 & 0 \\ 0 & h_2 v & 0 \\ 0 & 0 & h_3 v \end{array} \right),
\end{equation}
if $v_1=v_2=v_3=v$. This is the starting point of most subsequent A$_4$ 
models.  More recently, I discovered \cite{m06} that Eq.~(4) may also be 
obtained with $(\nu_i,l_i) \sim \underline{3}, ~l^c_i \sim \underline{3}$ and 
$(\phi^0_i,\phi^-_i) \sim \underline{1}, \underline{3}$, in which case
\begin{equation}
{\cal M}_l = \left( \begin{array}{ccc} h_0 v_0 & h_1 v_3 & h_2 v_2 \\ 
h_2 v_3 & h_0 v_0 & h_1 v_1 \\ h_1 v_2 & h_2 v_1 & h_0 v_0 \end{array} 
\right) = U_{l \nu}^{CW} \left( \begin{array}{ccc} m_e & 0 & 0 \\ 
0 & m_\mu & 0 \\ 0 & 0 & m_\tau \end{array} \right) (U_{l \nu}^{CW})^\dagger,
\end{equation}
if $v_1=v_2=v_3=v$.  Either way, $U_{l \nu}^{CW}$ has been derived.  To 
obtain $U_{l \nu}^{HPS}$, let ${\cal M}_\nu$ be Majorana and come from Higgs 
triplets: $(\xi^{++},\xi^+,\xi^0)$, then \cite{m04}
\begin{equation}
{\cal M}_\nu = \left( \begin{array}{ccc} a+b+c & f & e \\ f & a+\omega b + 
\omega^2 c & d \\ e & d & a + \omega^2 b + \omega c \end{array} \right),
\end{equation}
where $a$ comes from \underline{1}, $b$ from \underline{1}$'$, $c$ from 
\underline{1}$''$, and $(d,e,f)$ from \underline{3}.  To obtain Eq.~(5), 
we simply let $b=c$ and $e=f=0$.  Note that the tribimaximal mixing matrix 
does not depend on the neutrino mass eigenvalues $a-b+d$, $a+2b$, 
$-a+b+d$, nor the charged-lepton masses.  This implies the existence of 
residual symmetries exhibited by the mass matrices which allow one to 
reconstruct \cite{l07} the original symmetry of the Lagrangian.

Since \underline{1}$'$ and \underline{1}$''$ are unrelated in A$_4$, the 
condition $b=c$ is rather {\it ad hoc} .  A very clever solution was 
proposed by Altarelli and Feruglio \cite{af05}: they eliminated both 
\underline{1}$'$ and \underline{1}$''$ so that $b=c=0$. In that case, 
$m_1=a+d$, $m_2=a$, $m_3=-a+d$.  This is the simplest model of 
tribimaximal mixing, with the prediction of normal ordering of neutrino 
masses and the sum rule \cite{m05}
\begin{equation}
|m_{\nu_e}|^2 \simeq |m_{ee}|^2 + \Delta m^2_{atm}/9.
\end{equation}
Babu and He \cite{bh05} proposed instead to use 3 heavy neutral singlet 
fermions with ${\cal M}_D$ proportional to the identity and ${\cal M}_N$ 
of the form of Eq.~(5) with $b=0$.  In that case, the resulting 
${\cal M}_\nu$ has $b=c$ and $d^2=3b(b-a)$.  This scheme allows both 
normal and inverted ordering of neutrino masses.

The technical challenge in all such models is to break A$_4$ spontaneously 
along 2 incompatible directions: (1,1,1) with residual symmetry Z$_3$ and 
(1,0,0) with residual symmetry Z$_2$.  There is also a caveat.  If 
$\nu_2 = (\nu_e + \nu_\mu + \nu_\tau)/\sqrt{3}$ remains an eigenstate, 
i.e. $e=f=0$, but $b \neq c$ is allowed, then the bound $|U_{e3}| < 0.16$ 
implies \cite{m04} $0.5 < \tan^2 \theta_{12} < 0.52$, away from the 
preferred experimental value of $0.45 \pm 0.05$.

\section{Beyond A$_4$ [S$_4$, $\Delta$(27), $\Sigma$(81), Q(24)]}

The group of permutation of 4 objects is S$_4$.  It contains both S$_3$ 
and A$_4$.  However, since the \underline{1}$'$ and \underline{1}$''$ 
of A$_4$ are now combined into the \underline{2} of S$_4$, tribimaximal 
mixing is achieved only with Eq.~(9). Furthermore, $h_1 \neq h_2$ in 
${\cal M}_l$ now requires both \underline{3} and \underline{3}$'$ 
Higgs representations.  No advantage appears to have been gained.

The group $\Delta(27)$ has the interesting decomposition $\underline{3} \times 
\underline{3} = \underline{\bar{3}} + \underline{\bar{3}} + 
\underline{\bar{3}}$, which allows
\begin{equation}
{\cal M}_\nu = \left( \begin{array}{ccc} x & fz & fy \\ fz & y & fx \\ 
fy & fx & z \end{array} \right).
\end{equation}
Using $\tan^2 \theta = 0.45$ and $\Delta m^2_{atm} = 2.7 \times 10^{-3}$ 
eV$^2$, this implies \cite{m07-1} $m_{ee} = 0.14$ eV.

The subgroups $\Sigma(3n^3)$ of U(3) may also be of interest.  $\Sigma(81)$ 
has 17 irreducible representations and may be applicable \cite{m07-2} 
to the Koide lepton mass formula
\begin{equation}
m_e + m_\mu + m_\tau = (2/3)(\sqrt{m_e} + \sqrt{m_\mu} + \sqrt{m_\tau})^2,
\end{equation}
as well as neutrino tribimaximal mixing \cite{m07-3}.

Since A$_4$ is a subgroup of SO(3), it has a spinorial extension which is 
a subgroup of SU(2).  This is the binary tetrahedral group, which has 24 
elements with 7 irreducible representations: \underline{1}, \underline{1}$'$, 
\underline{1}$''$, \underline{2}, \underline{2}$'$, \underline{2}$''$, 
\underline{3}.  It is also isomorphic to the quaternion group Q(24) whose 
24 elements form the vertices of the self-dual hyperdiamond in 4 dimensions. 
There have been several recent studies \cite{fhlm07,cm07,fk07,a07} 
involving Q(24), which may be useful for extending the success of A$_4$ 
for leptons to the quark sector.

\section{Some Remarks}

With the application of the non-Abelian discrete symmetry A$_4$, a 
plausible theoretical understanding of tribimaximal neutrino mixing 
has been achieved. Other symmetries such as 
S$_4$, $\Delta$(27), $\Sigma$(81), and Q(24) are beginning to be studied. 
They share some of the properties of A$_4$ and may help to extend our 
understanding of possible discrete family symmetries, with eventual 
links to grand unification.

\section*{Acknowledgments}

I thank Xiaoyuan Li and Chun Liu for their great hospitality in Beijing 
and help at this very stimulating conference on flavor physics.  This 
work was supported in part by the U.~S.~Department of Energy under Grant 
No.~DE-FG03-94ER40837.


\begin{thebibliography}{00}

\bibitem{c78} N. Cabibbo, {\it Phys. Lett.} {\bf B72}, 333 (1978).
\bibitem{w78} L. Wolfenstein, {\it Phys. Rev.} {\bf D18}, 958 (1978).
\bibitem{hps02} P. F. Harrison, D. H. Perkins, and W. G. Scott, 
{\it Phys. Lett.} {\bf B530}, 167 (2002).
\bibitem{m04} E. Ma, {\it Phys. Rev.} {\bf D70}, 031901 (2004).
\bibitem{m02} E. Ma, {\it Mod. Phys. Lett.} {\bf A17}, 2361 (2002).
\bibitem{mr01} E. Ma and G. Rajasekaran, {\it Phys. Rev.} {\bf D64}, 113012 
(2001).
\bibitem{bmv03} K. S. Babu, E. Ma, and J. W. F. Valle, {\it Phys. Lett.} 
{\bf B552}, 207 (2003).
\bibitem{m06} E. Ma, {\it Mod. Phys. Lett.} {\bf A21}, 2931 (2006).
\bibitem{l07} C. S. Lam, {\it these proceedings.}
\bibitem{af05} G. Altarelli and F. Feruglio, {\it Nucl. Phys.} {\bf B72)}, 
64 (2005).
\bibitem{m05} E. Ma, {\it Phys. Rev.} {\bf D72}, 037301 (2005).
\bibitem{bh05} K. S. Babu and X.-G. He, hep-ph/0507217.
\bibitem{m07-1} E. Ma, arXiv:0709.0507 [hep-ph].
\bibitem{m07-2} E. Ma, {\it Phys. Lett.} {\bf B649}, 287 (2007).
\bibitem{m07-3} E. Ma, {\it Eur. Phys. Lett.} {\bf 79}, 61001 (2007).
\bibitem{fhlm07} F. Feruglio, C. Hagedorn, Y. Lin, and L. Merlo, {\it Nucl. 
Phys.} {\bf B775}, 120 (2007).
\bibitem{cm07} M.-C. Chen and K. T. Mahanthappa, {\it Phys. Lett} {\bf B652}, 
34 (2007).
\bibitem{fk07} P. H. Frampton and T. W. Kephart, {\it JHEP} {\bf 0709}, 110 
(2007).
\bibitem{a07} A. Aranda, arXiv:0707.3661 [hep-ph].


\end{thebibliography}
\end{document}